\documentclass[5p,twocolumn,authoryear]{elsarticle}
\usepackage{hyperref}
\pdfoutput=1

\usepackage{multirow}
\usepackage{multicol}
\usepackage{color}
\usepackage{amsmath}
\usepackage{amssymb}
\allowdisplaybreaks
\usepackage{caption}
\usepackage{subcaption}
\usepackage{epstopdf}

\newcommand{\mycite}{\citep}
\renewcommand{\vec}[1]{\mathbf{#1}}

\graphicspath{ {./figures/} }

\begin{document}

\begin{frontmatter}
\title{Drag correlation for dilute and moderately dense fluid-particle systems using the lattice Boltzmann method}

\author[LSS]{Simon Bogner\corref{cor}}
\ead{simon.bogner@fau.de}

\author[IMMT]{Swati Mohanty\corref{cor}}
\ead{swati.mohanty@gmail.com}

\author[LSS]{Ulrich R\"ude}
\ead{ruede@fau.de}

\address[IMMT]{CSIR-Institute of Minerals and Materials Technology, Bhubaneswar-751013, India}

\address[LSS]{Lehrstuhl f\"ur Systemsimulation, 
	Universit\"at Erlangen-N\"urnberg,
	Cauerstra{\ss}e 11,
	91058 Erlangen,
        Germany}

 \cortext[cor]{Corresponding author}

\begin{abstract}
  This paper presents a numerical study of flow through static random assemblies of
monodisperse, spherical particles. A lattice Boltzmann approach based on a two relaxation
time collision operator is used to obtain reliable predictions of the particle drag by
direct numerical simulation. From these predictions a closure law $F(Re_p,\varphi)$ of the
drag force relationship to the bed density $\varphi$ and the particle Reynolds number
$Re_p$ is derived.
The present study includes densities $\varphi$ ranging from $0.01$ to $0.35$ with $Re_p$
ranging up to $300$, that is compiled into a single drag correlation valid for the whole
range. The corelation has a more compact expression compared to others previously
reported in literature. 
At low particle densities, the new correlation is close to the widely-used Wen \& Yu
-- correlation.

Recently, there has been reported a discrepancy between results obtained using different
numerical methods, namely the comprehensive lattice Boltzmann study of
\citet{Beetstra2007} and the predictions based on an immersed boundary - pseudo-spectral
Navier-Stokes approach \mycite{Tenneti2011}. The present study excludes significant finite
resolution effects, which have been suspected to cause the reported deviations, but does
not coincide exactly with either of the previous studies. This indicates the need for yet
more accurate simulation methods in the future.


\end{abstract}


\end{frontmatter}

\section{Introduction}

Many of the mineral processing unit operations involve multi-phase systems that can be
classified as either solid-liquid system such as hydrocyclone, high wet intensity magnetic
separator, fluidized bed, thickener, spiral concentrator; solid-gas system such as gas
fluidized bed; or solid-gas-liquid system such as flotation column. The volume fraction of
solid particles that is fed into these units varies from sometimes very low concentrations
as in the case of hydrocyclones or magnetic separators to very high concentrations as in
the case of fluidized beds. The mineral processing industry is considered as one of the
most polluting industries and therefore it is desirable to use only as few experiments as
possible. Computer simulations prior to and accompanying experiments cannot only reduce
the number of experiments, but are also cheaper. With the advent of high speed computers,
it is now possible to obtain solutions faster and with better accuracy. This does not only
save time, money and material, but is also pollution-free.

Several approaches like continuum models \mycite{Swain2013, Mohanty2011}, discrete models
\mycite{Mishra2010} and a combination of discrete and continuum models \mycite {Chu2009}
have been reported in literature for various mineral processing unit operations. In
practice, multi-fluid models such as Eulerian-Eulerian models, which treat the phases as
inter-penetrating continua, are more common because in an actual system it is difficult to
capture the dynamics of all individual particles present. It has been reported by
\citet{Hoeff2006} that for millimeter size particles the linear dimension of the system
that can be simulated with the existing computers is about $0.1$ m. For larger systems,
multi-fluid continuum models have become more popular that treat the granular solid phase
as a continuum and the momentum balance equations for both the phases are solved to
predict the flow profile of the phases. These mathematical models require several closure
laws, primarily for the drag force, to account for the momentum exchange between the
phases. The earliest correlations available were those by \citet{Wen1966} for dilute
systems and the Ergun's equation \mycite{Ergun1952} for denser systems. Both are based on
experimental data. Other correlations commonly used in two-fluid models are the ones
proposed by \citet{Gidaspow1986} and \citet {Shyamlal1987}. The Gidaspow correlation
reduces to the Wen and Yu correlation for low solid volume fraction and Ergun's equation
at high solid volume fraction. The correlation by Shyamlal and O'Brien is based on the
terminal velocity correlation proposed by \citet{Richardson1954}. The reliability of these
correlations for mono-dispersed system has been questioned by some authors \mycite
{Beetstra2007}, as Ergun derived the correlation based on experiments carried out using
sand and pulverized coke in addition to that carried out with spheres.

\subsection{Studies on drag closure laws based on DNS}
\label{sec:intro:dns}
With the advent of high speed computers, several direct numerical simulation (DNS) methods
have been adopted by a number of researchers to obtain drag closure laws. Pioneering
studies on particulate suspensions by \citet{Ladd1994a, Ladd1994b, Ladd2001} using the
lattice Boltzmann method (LBM) enabled the work on establishing drag closure laws. The
earliest systematic studies on drag force relations by \citet{Hill2001a} considered a
fixed bed of spheres arranged in simple cubic, face centered and random order. The volume
fraction of solids ranged from approximately $0.001$ to $0.64$ at low Reynolds
number $Re_p$. Subsequently, the effect of the Reynolds number, up to $Re_p \leq 100$, on the
drag force for the same arrangement of spheres was studied by the same authors
\mycite{Hill2001b}. Later, \citet{vanDerHoef2005} used LBM to study the drag force on
fixed arrays of mono-dispersed and bi-dispersed spheres arranged in random order in the
limit of zero Reynolds number. They established a normalized drag force correlation,
defined as the ratio of the drag force for a given Reynolds number and solid volume
fraction to that calculated using the Stokes-Einstein equation. They also reported that
when their drag force correlation was extrapolated to $Re_p=0$, there was no significant
difference between drag force at $Re_p=0.2$ and $Re_p=0$. \citet{Beetstra2007} later
extended this work to $Re_p$ up to $1000$. They also obtained a correlation for the
normalized drag force as a function of solid volume fraction and Reynolds
number. Reportedly, the Ergun type correlation fits the simulation data better than the
Wen and Yu type. \citet{Benyahia2006} modified the correlations proposed by
\citet{Hill2001a} so that a continuous function can be obtained for all range of solid
volume fraction and Reynolds number.  \citet{Tenneti2011} have used a particle resolved
DNS method that they call ``Particle-resolved Uncontaminated-fluid Reconcilable Immersed
Boundary Method'' to study the drag force on a fixed array of randomly arranged assembly
of mono-dispersed spheres in the solid volume fraction range of 0.1-0.4 and Reynolds
number up to 300. The results are compared with \citet{Hill2001b} and
\citet{Beetstra2007}. The three studies agree quite well at low Reynolds number. At higher
Reynolds number, the data of \citet{Tenneti2011} and \citet {Beetstra2007} disagree by up
to $30 \%$. \citet{Tenneti2011} attribute the mismatch to the constant resolution used in
\citet{Beetstra2007} over the whole range of Reynolds numbers.

All of the above studies assume static beds of spheres without changes in the relative
positions. However, some work on moving particles for bi-dispersed spheres of equal size
but different densities \mycite{Yin2009a} as well as of different sizes \mycite{Yin2009b}
at low Reynolds number and solid volume fraction of $0.1 .. 0.4$, have been reported using
LBM. \citet{Holloway2010} used the data from \citet{Yin2009a,Yin2009b} and that by
\citet{Beetstra2007} to obtain a drag correlation for poly-dispersed particles in the
solid volume fraction range $0.2 .. 0.4$, particle diameter ratio between $1$ and $2.5$
and $0 \leq Re_p \leq 40$.

\subsection{Objective of the present study}
\label{sec:intro:objective}
Most studies using LBM have reported drag laws for solid volume fraction above $0.1$ and higher
Reynolds numbers, primarily aimed at fluidized beds. Hence, there is currently a lack of
studies carried out in the low solid volume fraction range, except for the results reported
by \citet{Hill2001b} for low Reynolds numbers. Several unit operations in mineral
processing operate with lower solid volume fractions and higher Reynolds number. Also, in
the same system there can be regions with low as well as high solid
volume fraction. Therefore a drag law applicable over a wider range of particle concentration, including
solid volume fraction $\leq 0.1$ at a given range of Reynolds number would be desirable.
In the following study, a drag law for solid-fluid dispersions with solid volume fraction
ranging from $0.01$ to $0.35$ and Reynolds numbers $Re_p \leq 300$ is derived from
simulations.

All the lattice Boltzmann studies on fluid-particle drag mentioned in
Sec.~\ref{sec:intro:dns} have been carried out with the code developed by Ladd
\mycite{Ladd1994a, Ladd1994b, Ladd2001}. One common disadvantage of these efforts is that
the sphere is represented by a staircase like approximation (\emph{bounce back} boundary
conditions) in simulations and thus a certain error will be introduced. However, it has
been established already in the original work that LBM still delivers accurate
predictions, for instance by comparing it with analytical solutions for Stokes flow of
\citet{Hasimoto1959} and \citet{Sangani1982}. Another common disadvantage of the
aforementioned LBM studies is the usage of a single relaxation time collision
operator. This is known to introduce a viscosity-dependent error in the boundary placement
\mycite{Ginzburg1994}, and forces the authors to apply a correction for the effective
hydrodynamic radius of the particles for the given relaxation time. The present work is
based on a two relaxation time lattice Boltzmann code, tailored to circumvent the
viscosity-dependent boundary placement \mycite{Ginzburg1994,Ginzburg2007} to obtain more
reliable results.  Details on the simulation methodology are given in
Sec.~\ref{sec:method}. Simulation results can be found in Sec.~\ref{sec:results}. The
presentation of the main results of this paper in
Sec.~\ref{sec:qualitative}~-~\ref{sec:comparison} is preceeded by a verification
experiment of the numerical model (Sec.~\ref{sec:regular}). Most importantly, we have
conducted a resolution sensitivity study to show that there is no significant error
stemming from under-resolved physics in the present study (Sec.~\ref{sec:resolution}).

Without relying on supercomputers the present study would not have been successful, as
simulation of dilute particle concentrations at high Reynolds numbers is quite resource
demanding. E.g., for $\varphi=0.01$ at a resolution of $d=40$ lattice sites per particle
diameter, one needs a domain size of approximately $450^3$ grid points. To accomplish
efficient simulations, the present study is based on the waLBerla code base
\citep{walberla2011}. This framework has been shown to offer extendibility to numerous
physical applications involving fluid-structure interaction \citep[e.g.,][]{Bogner2013},
while providing a highly efficient implementation for a wide range of supercomputers, even
for complex fluids like suspensions \citep{Goetz2010}.


\section{Numerical Method: Two Relaxation Time LBM for Particle Beds}
\label{sec:method}For the simulated hydrodynamics we employ a lattice Boltzmann approach
\mycite{BenziEtAl,ChenDoolen1998,AidunClausen} using the \emph{D3Q19} lattice model
\mycite{QianEtAl1992}. The evolution of the distribution function $\vec{f} = (f_0, f_1,
.., f_{Q-1})$ on the lattice for the finite set of \emph{lattice velocities} $\{ \vec{c}_q
\, | \, q=0, .., Q-1 \}$ is described by the following equations, Eq.s (\ref{eq:lbe1}) and
(\ref{eq:lbe2}). We use a collision operator with two relaxation times as proposed by
\citet{Ginzburg2007,Ginzburg2005}.
\begin{equation}
  f_q (\vec{x}+\vec{c}_q, t+1) = \tilde{f_q} (\vec{x}, t),\label{eq:lbe1} \\
\end{equation}
\begin{equation}
  \tilde{f_q} (\vec{x}, t) = f_q (\vec{x}, t) + \lambda_{+}(f_q^{+} - f_q^{eq,+}) + \lambda_{-}(f_q^{-} - f_q^{eq,-}) + F_q,
  \label{eq:lbe2}
\end{equation}
Here, Eq.~(\ref{eq:lbe1}) is referred to as the \emph{streaming step}, and
Eq.~(\ref{eq:lbe2}) is the \emph{collision step}, with independent relaxation times
$\lambda_{+}, \lambda_{-} \in (-2, 0)$ for the \emph{even (symmetric)} and \emph{odd
  (anti-symmetric)} parts of the distribution function. Often the symmetric relaxation
time is expressed as $\tau = -1 / \lambda_{+} \in (0.5, \infty)$. $F_q$ is a source term
that will be defined later. For the discrete range of values $q \in \{0, .., Q-1\}$, the
opposite index $\bar{q}$ is defined by the equation $\vec{c}_q = - \vec{c}_{\bar{q}}$ and
thus,
\begin{align}
  f_q^{+} = \frac{1}{2}(f_q + f_{\bar{q}}),\;\text{ and }\; f_q^{-} = \frac{1}{2}(f_q - f_{\bar{q}}),
\end{align}
respectively.  The polynomial equilibrium function $\vec{f}^{eq}$ is given by
\begin{equation}
  f_{q}^{eq}( \rho, \vec{U}) = \rho w_{q} \left[ 1 + \frac{ \vec{c}_q^T \vec{U} }{c_s^2} + \frac{ (\vec{c}_q^T \vec{U})^2 }{2 c_s^4} - \frac{ \vec{U}^T \vec{U} }{2 c_s^2} \right],
\end{equation}
where the $w_q = w_{|\vec{c}_q|}$ are the \emph{lattice weights} given in
\citet{QianEtAl1992}, and $c_s= 1/\sqrt{3}$ is the \emph{lattice speed of sound} of the
model. To exert a body force on the flow we use the source term
\begin{equation}
  F_q = w_q \rho \left( \frac{\vec{c}_q-\vec{U}}{c_s^2} + \frac{\vec{c}_q^T \vec{U}}{2 c_s^4} \vec{c}_q \right)^{T} \vec{a},
\end{equation}
where $\vec{a}$ is a gravitation vector \mycite{Luo1998,Guo2002}. Macroscopic quantities
are defined as moments over $\vec{f}$. In particular, the moments of zeroth and first
order,
\begin{align} 
  \rho = \frac{1}{c_s^2} P &= \sum_q f_q,
  \label{eq:density} \\
  \rho \vec{U} = \rho \vec{u} - \frac{\rho \vec{a}}{2} &= \sum_q \vec{c}_q f_q,
  \label{eq:momentum}
\end{align}
yield the pressure $P$ and fluid velocity $\vec{u}$. The shift in the fluid momentum by
$\rho \vec{a}/{2}$ is necessary if a source term $F_q$ is used in Eq. (\ref{eq:lbe2})
\citep[cf.][]{BuickGreated2000,Ginzburg2007}. The parameterization of $f^{eq}$ in
Eq. (\ref{eq:lbe2}) stays unchanged and is given by $f^{eq}_q = f^{eq}_q(\rho, \vec{U})$.

\paragraph{Fluid-structure interaction}
To incorporate spherical particles into the flow, we use the bounce-back rule similar to
\cite{Ladd1994a,Ladd1994b}, for all lattice links intersecting with the surface of a
particle. For a boundary node $\vec{x}_b$ with boundary-intersecting link $q$ one has
\begin{equation}
  f_{\bar{q}}(\vec{x}_b, t+1) = \tilde{f_q} (\vec{x}_b, t).
\end{equation}
This approximation of a non-slip boundary condition can be made effectively
independent of the lattice viscosity, $\eta = -(\lambda_{+}^{-1}+1/2)/3 = (\tau -1/2)/3$ by fixing the
second relaxation time $\lambda_{-}$, to satisfy
\begin{equation}
  \Lambda_{\pm} := (\frac{1}{2} + \frac{1}{\lambda_{+}})(\frac{1}{2} + \frac{1}{\lambda_{-}}) \approx \frac{1}{4},
  \label{eq:magic}
\end{equation}
as shown by \citet{Ginzburg1994,Ginzburg2007,Ginzburg2009}. The optimal value of
$\lambda_{-}$ may slightly differ depending on the geometry of the flow
\mycite{Ginzburg2007}. Here, we used $\Lambda_{\pm} = 3/16$. This is an important
difference of the present approach from previous lattice Boltzmann studies on drag
relations such as those reported by \citet{vanDerHoef2005}, \citet{Beetstra2007} and
\citet{Hill2001a,Hill2001b}. These studies are all based on collision schemes that do not
allow viscosity-independent simulations of flow around particles and thus come with an
additional source of error, which can be controlled only by fixing the lattice viscosity
to a constant value. \citet{Pan2005} have shown that the elimination of the
viscosity-dependent error is significant, especially when relaxation times $\tau>1.0$ are
used, as typically the case for low Reynolds-number flows. It should be noted, that the
bounce-back rule yields a first-order accurate approximation of the particle boundaries
\citep{Ginzburg2007,JunkYang2005}. Higher-order schemes
\citep{Bouzidi2001,Ginzburg2003,MeiYuShyyLuo2002} exist, but are more complex to apply
since they require interpolations over a number of compute nodes. Also, according to
\citep{Ginzburg2007,Ginzburg2009,Pan2005}, not all of them allow viscosity-independent
parameterization according to Eq.~(\ref{eq:magic}), and thus may complicate the
parameterization in terms of lattice viscosity. Hence, for the present study, we decided
to use only the bounce-back rule, but to include a resolution sensitivity study, showing
that the error due to boundary conditions is sufficiently converged
(cf. Sec.~\ref{sec:resolution}).

The drag force exerted by the fluid on the particles is obtained by the \emph{momentum
  exchange algorithm} \mycite{Ladd1994a,Yu2003}. For a particle $p$, let $B_p$ be the set
of boundary nodes, where for all $\vec{x_b} \in B_p$, the set of links $I_p(\vec{x_b})
\subseteq \{0,..,Q-1\}$, intersecting with the surface of the particle $p$ is non-empty. The
net force $\vec{f}_p$ exerted on the particle is then given by
\begin{equation}
  \vec{f}_p = \sum_{\vec{x}_b \in B_p} \sum_{q \in I_p(\vec{x}_p)} \Delta \vec{j}_q(\vec{x}_b),
  \label{eq:mea}
\end{equation}
where $\Delta \vec{j}_q(\vec{x}_b) = \vec{c}_q (\tilde{f}_q + f_{\bar{q}})$ is the
momentum transferred along a single intersecting link in direction $q$ from the boundary
node $\vec{x}_b$. Note that in our specific case, the incoming population $f_{\bar{q}}$
is actually equal to the outgoing $\tilde{f}_q$ because of the bounce-back rule.

\paragraph{Drag force computations}
The flow through random fixed configurations (beds) of $N$ non-overlapping
spherical particles of diameter $d$ is simulated by accelerating the flow along a certain direction
with a given uniform gravity $g$. The drag force evaluation is done after the flow has
reached a balanced state. The particle Reynolds number is expressed as,
\begin{equation}
  Re_p = \frac{\rho \bar{u} d}{\eta},
\end{equation}
where
\begin{equation}
  \bar{u} = \frac{\vec{a}^T}{|\vec{a}| L^3} \sum_{\vec{x} \in \Omega} (\vec{u}(\vec{x}) - \vec{u}_p)
\end{equation}
is the \emph{average flow rate} (directed with gravity $\vec{a}$) within a periodic cubic
domain $\Omega$ of length $L$, relative to the particle velocity $\vec{u}_p$, that is
taken as zero, since the particle positions are fixed with respect to the given frame of
reference. The characteristic length $d$ can also be interpreted as the Sauter mean
diameter that is often used to study heterogeneous, non-spherical particle beds. The
approximated drag force $\vec{f}_d$ is obtained by averaging Eq. (\ref{eq:mea}) over the
whole number of particles, i.e., $\vec{f}_d = \frac{1}{N} \sum_{p=1}^{N} \vec{f}_p$. The
solid volume fraction of the bed is given by
\begin{equation}
  \varphi = \frac{N V_p}{V},
\end{equation}
where $V_p$ is the particle volume, and $V=L^3$ the total volume of the bed. 

In many practically relevant conditions, the flow in the particle bed will be driven by a
static pressure gradient, yielding an additional buoyancy force $\vec{f}_b = -V_p \nabla
P$ on each particle. Hence, the total hydrodynamic force (averaged per particle) is
actually $\vec{f}_t = \vec{f}_d + \vec{f}_b$. Often, $\vec{f}_t$ is referred to as the drag
force in literature. However, assuming the system to be in a balanced state, i.e., the
driving force equals the total force exerted on the particles ($-V \nabla P = N \vec{f}_t$), the
two definitions of the drag force can be directly related as $\vec{f}_d = (1-\varphi)
\vec{f}_t$. Since a pressure gradient cannot be included in simulations using periodic
domains, we will simply assume $\vec{f}_b = V_p \rho \vec{a}$ and interpret the results as
equivalent to a pressure driven flow. The \emph{dimensionless drag force} is accordingly
defined as
\begin{equation}
  C = \frac{ f_t }{ 3 \pi \eta d \bar{u} },\text{ or  } C_d = \frac{ f_d }{ 3 \pi \eta d \bar{u} },
\end{equation}
respectively.


\section{Results: Simulation of Flow in Random Particle Beds}
\label{sec:results}After two verification studies in Sec.~\ref{sec:regular} and
Sec.~\ref{sec:resolution}, demonstrating the validity of our method, we study the flow
through randomly generated fixed arrangements of spheres. Sec.~\ref{sec:qualitative}
presents qualitative results for the most extreme cases studied. Finally, we present a
comprehensive and systematic flow study in Sec.~\ref{sec:comprehensive} including its
compilation into a drag force closure relation and compare the result with two other
numerical studies in Sec.~\ref{sec:comparison}. Quantities obtained as results from
numerical experiments will be labeled with a superscript $*$ in the following.

\subsection{Verification: Stokes flow in regular simple cubic array}
\label{sec:regular}
For the verification of our code, we first evaluate the linear flow through a regular
periodic array of spheres. For the case of a \emph{simple cubic array},
\citet{Sangani1982} have presented analytical results valid over a wide range of solid
volume fractions $\varphi(\chi) = \pi \chi^3 /6$, extending the prior work of
\citet{Hasimoto1959}. By varying the radius $R$ of a sphere placed within a periodic unit
cell of length $L=32$ and $L=64$ in lattice units, respectively, a set of different
$\varphi$ is realized. The normalized drag force obtained from simulations, $C^*$, and the
error relative to the analytical solution $C$ by \citet{Sangani1982} is shown in
Tab.~\ref{table:drag}. The average flow rate $\bar{u}^*$ in these simulations is also
shown. Note, that the Mach and Reynolds numbers are always kept low (i.e., $u / c_s <
0.01$ and $Re_p < 0.03$) for exclusion of compressibility effects and assuring a good
approximation to the Stokes regime, respectively. To speed up the convergence, the Mach
number could be increased (typically one chooses $0.01 < u/c_s < 0.1$ \citep{Succi} - here
it was easier to compute with constant $\lambda_{+} = -1/3$ and gravitation
$10^{-6}$). The relative errors are below $5\%$ in all cases, which shows the validity of
our code. It also shows that the method is capable of accurate predictions of the drag
force on spheres, even if the resolution of the sphere, or the gap in between the spheres,
respectively, is relatively small.

\begin{table*}
\centering
\small
\setlength{\tabcolsep}{0.12cm}
\begin{tabular}{|l| *{9}c| }
\hline
\multicolumn{10}{ |c| }{Drag Force Computation in Simple Cubic Array} \\
\hline
& $\varphi$ &  0.0042 & 0.014 & 0.034 & 0.065 & 0.11 & 0.18 & 0.27 & 0.38 \\
$L$& $\chi$ & 0.2 & 0.3 & 0.4 & 0.5 & 0.6 & 0.7 & 0.8 & 0.9 \\ 
\cline{2-10}

\multirow{4}{*}{$32$} 
& R &  3.2 & 4.8 & 6.4 & 8 & 9.6 & 11.2 & 12.8 & 14.4 \\
& $\bar{u}^*$ &   4.75E-4 & 2.49E-4 & 1.51E-4 & 8.91E-5 & 5.45E-5 & 3.03E-5 & 1.59E-5 & 7.21E-6 \\
& $C^*$ &  1.37 & 1.75 & 2.16 & 2.93 & 4.00 & 6.15 & 10.29 & 20.06 \\
& $\frac{C^*-C}{C}$ &  -1.13\% & 2.74\% & 0.45\% & 2.93\% & 0.57\% & 2.49\% & 2.43\% & 4.69\% \\
\cline{2-10}

\multirow{4}{*}{$64$} 
& R & 6.4 & 9.6 & 12.8 & 16 & 19.2 & 22.4 & 25.6 & 28.8 \\
& $\bar{u}^*$ &  1.87E-3 & 1.02E-3 & 6.00E-4 & 3.61E-4 & 2.17E-4 & 1.23E-4 & 6.36E-5 & 2.97E-5 \\
& $C^*$ &  1.39 & 1.70 & 2.17 & 2.89 & 4.00 & 6.08 & 10.25 & 19.52 \\
& $\frac{C^*-C}{C}$ &  0.37\% & 0.27\% & 0.92\% & 1.63\% & 0.73\% & 1.27\% & 1.95\% & 1.86\% \\
\hline
\end{tabular}
\caption{Drag force on spheres in simple cubic array and the relative error compared to the analytic solution from \citet{Sangani1982} at two different resolutions. }
\label{table:drag}
\end{table*}

\subsection{Verification: Resolution sensitivity}
\label{sec:resolution}

\begin{table}
  \centering

    \small  
    \setlength{\tabcolsep}{0.18cm}
    \begin{tabular}{|l| *{5}{c|} }
      \hline
      \multicolumn{6}{ |c| }{(a) $Re_p = 100$} \\
      \hline
      $\delta_x$& $\tau$ & $d / \delta_x$ & $Re_p^{*}$ & $C_d^{*}$ & $\delta_b/\delta_x$ \\ 
      \hline 
      0.02 (!) &  0.500563 & 14.5726 & 102 & 26.6 & 1.4572 \\
      0.015625 &  0.500922 & 18.653 & 100 & 27.2 & 1.8653 \\
      0.01     &  0.50225  & 27.5089 & 99  & 27.5  & 2.751 \\
      0.005    &  0.509    & 58.2906 & 100 &  27.2  & 5.8291 \\
      \hline

      \multicolumn{6}{ c }{ }\\ 

      \hline
      \multicolumn{6}{ |c| }{(b) $Re_p = 200$} \\
      \hline
      $\delta_x$& $\tau$ & $d / \delta_x$ & $Re_p^{*}$ & $C_d^{*}$ & $\delta_b/\delta_x$ \\ 
      \hline 
      0.015625 (!) & 0.500922 & 18.653  & 202 & 38.5& 1.319 \\
      0.01         & 0.50225  & 29.1453 & 198 & 41.3& 2.0609 \\
      0.005        & 0.509    & 58.2906 & 198 & 41.5& 4.1218 \\
      0.0025       & 0.536    & 116.581 & 202 & 40.5& 8.2435 \\
      \hline

      \multicolumn{6}{ c }{ }\\ 

      \hline
      \multicolumn{6}{ |c| }{(c) $Re_p = 300$} \\
      \hline
      $\delta_x$& $\tau$ & $d / \delta_x$ & $Re_p^{*}$ & $C_d^{*}$ & $\delta_b/\delta_x$ \\ 
      \hline 
      0.00625 & 0.50576 & 46.6325 & 300 & 55.4 & 2.6923 \\
      0.005   & 0.509   & 58.2906 & 300 & 55.6 & 3.3654 \\
      0.0025  & 0.536   & 116.581 & 300 & 54.0 & 6.7308 \\
      \hline
    \end{tabular}
    \caption{Sensitivity of drag coefficient on resolution. The cases marked with (!) would eventually become unstable. Also shown is the approximate boundary layer thickness $\delta_b$ in lattice units.}
    \label{table:resolutionRe100}
    \label{table:resolutionRe200}
    \label{table:resolutionRe300}


\end{table}

\begin{table}
  \centering
  
  \small  
  \setlength{\tabcolsep}{0.18cm}
  \begin{tabular}{|l c| *{3}{c|} }
    \hline
    & & \multicolumn{3}{ |c| }{$d/\delta_x$} \\
    & & 46.6325 & 58.2906 & 116.581 \\
    \hline
    \multirow{3}{*}{$\Lambda_{\pm}$} 
    & 3/16 & 55.4 & 55.6 & 54.0 \\
    & 1/4  & 55.0 & 54.0 & 53.5 \\
    & 3/8  & 53.4 (!) & 54.0 & 53.4 \\
    \hline
  \end{tabular}
  \caption{Normalized drag $C_d^*$ for various ``magic parameter'' settings in the $Re_p=300$ -- case (c).}
  \label{tab:magic}

\end{table}

For particle Reynolds numbers in the unsteady regime numerical instabilities may arise in
under-resolved cases. We therefore study the dependency of the estimated drag $C_d^{*}$
obtained from simulations on the grid spacing $\delta_x$. Our setup consists of a
randomly arranged bed of $N=27$ spheres in a cubic periodic domain of unit length and
solid volume fraction $\varphi=0.35$. In order to keep the Reynolds number constant while
varying the resolution parameter $\delta_x$, we adjust the sphere diameter and lattice
relaxation time accordingly (i.e., the apparent physical viscosity and length is kept
constant).  The study is repeated for three different target Reynolds numbers $Re_p=100$,
$200$ and $300$.

For the first two cases ($Re_p=100$, $200$), the flow is driven by a constant body force
starting from zero uniform initial velocity, and evaluated after a balanced state is
reached. Tab.~\ref{table:resolutionRe100} shows the results for different Reynolds
numbers. Due to the ``staircase'' discretization of the spheres the resulting $Re_p^{*}$
varies slightly. For the highest Reynolds number, $Re_p=300$, we dynamically adjusted the
fluid acceleration $\vec{a}$ during simulations, in order to circumvent a manual
calibration process. Even in the under-resolved situations marked with an (!)  where one
would expect instability, we observe that the predicted $C_d^{*}$-value does not deviate
significantly from the well-resolved cases. From the magnitude of the deviations in the
data (less than $3\%$ when excluding the unstable solutions), it can be concluded that the
chosen lattice resolutions are sufficiently high. In fact, the results show that there is
a minimum required resolution that increases with the Reynolds number, but also that
further increase of resolution does not significantly effect the resulting drag
forces. This is an important observation, since it has been reported by
\citet{Tenneti2011} that previous studies based on the LBM \mycite{Beetstra2007} would
lack validity at intermediate Reynolds numbers due to poor lattice resolutions. As noted
by \citet{Tenneti2011}, the boundary layer thickness ($\delta_b \sim d/\sqrt{Re_p}$) must
be sufficiently resolved and can be used as a guideline when choosing the spatial
resolution.  For the given regime, one should have $\delta_b \sim 3$ or higher for numeric
stability. Further increase of resolution yields only slight improvement in terms of
accuracy.  Hence for the following study in
Sec.s~\ref{sec:qualitative}-\ref{sec:comprehensive}, we chose grid spacings of the order
of the second-to-last row of Tab.~\ref{table:resolutionRe100} as a compromise between
accuracy and efficiency.

\paragraph{Influence of parameterization $\Lambda_{\pm}$ on numerical solution}

For completeness, we also studied the influence of the magic parameter $\Lambda_{\pm}$ on
the simulations in the transitional regime $Re_p=300$. This allows us to estimate the
errors stemming from the parameterization of our scheme in Sec.~\ref{sec:comparison}, when
comparing to the results from other authors -- and the variation between the respective
studies is most considerable for large $Re_p$. The optimal choice of $\Lambda_{\pm}$ is
unknown for arbitrary geometries, so it is important to control whether the computed
solutions depend on it. Because, for this regime, one has only small values of $\tau$, the
effect on the solution is expected to be non-significant (cf. also \citet{Pan2005}). In
Tab.~\ref{tab:magic} the resulting drags for different parameterizations is
shown. Indeed, the influence was very small for the given geometry at the given
relaxation times $\tau$. The parameterization $\Lambda_{\pm}=3/8$ lead to instabilities at
the lowest resolution.

\subsection{Study: Qualitative evaluation of different flow regimes}
\label{sec:qualitative}

\begin{figure*}
\includegraphics[width=0.5\linewidth]{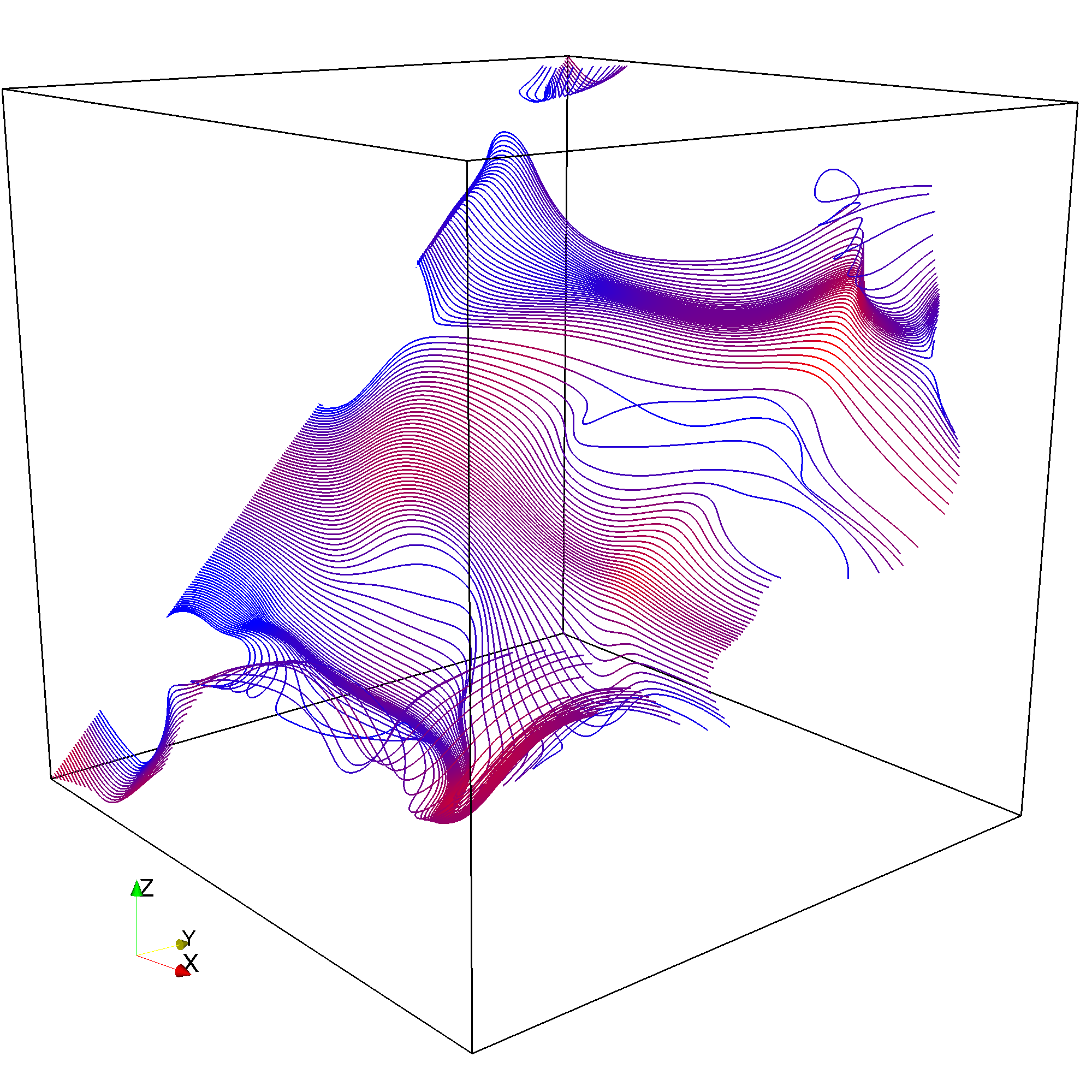}
\includegraphics[width=0.5\linewidth]{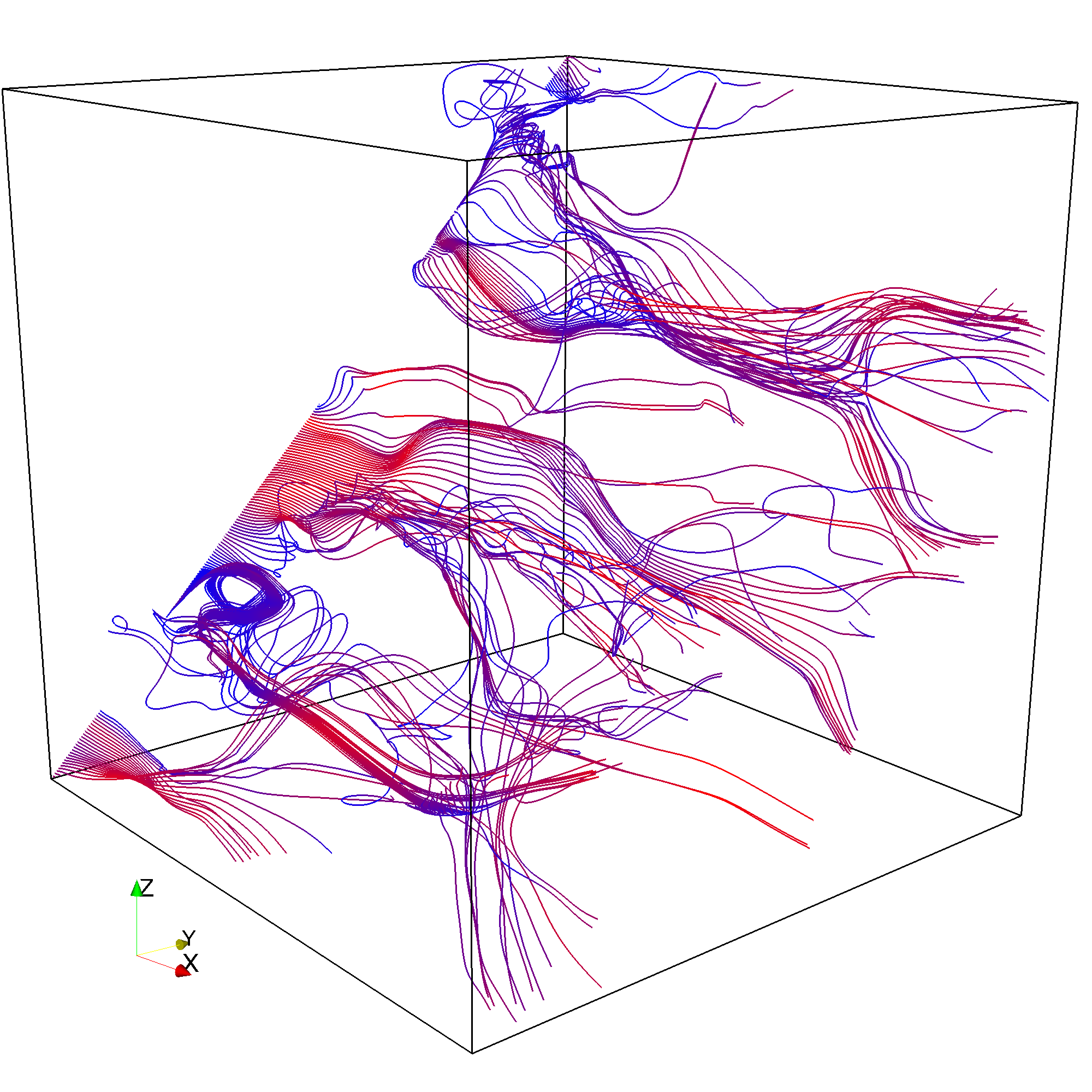}
\caption{Streamlines in bed of solid volume fraction $\varphi=0.35$ in the laminar ($Re_p =
  3.6$, left), and in the transient regime ($Re_p = 300$, right). Blue color indicates lower,
  red color indicates a higher flow velocity, respectively.}
\label{fig:streamlines}
\end{figure*}
\begin{figure*}
\includegraphics[width=0.5\linewidth]{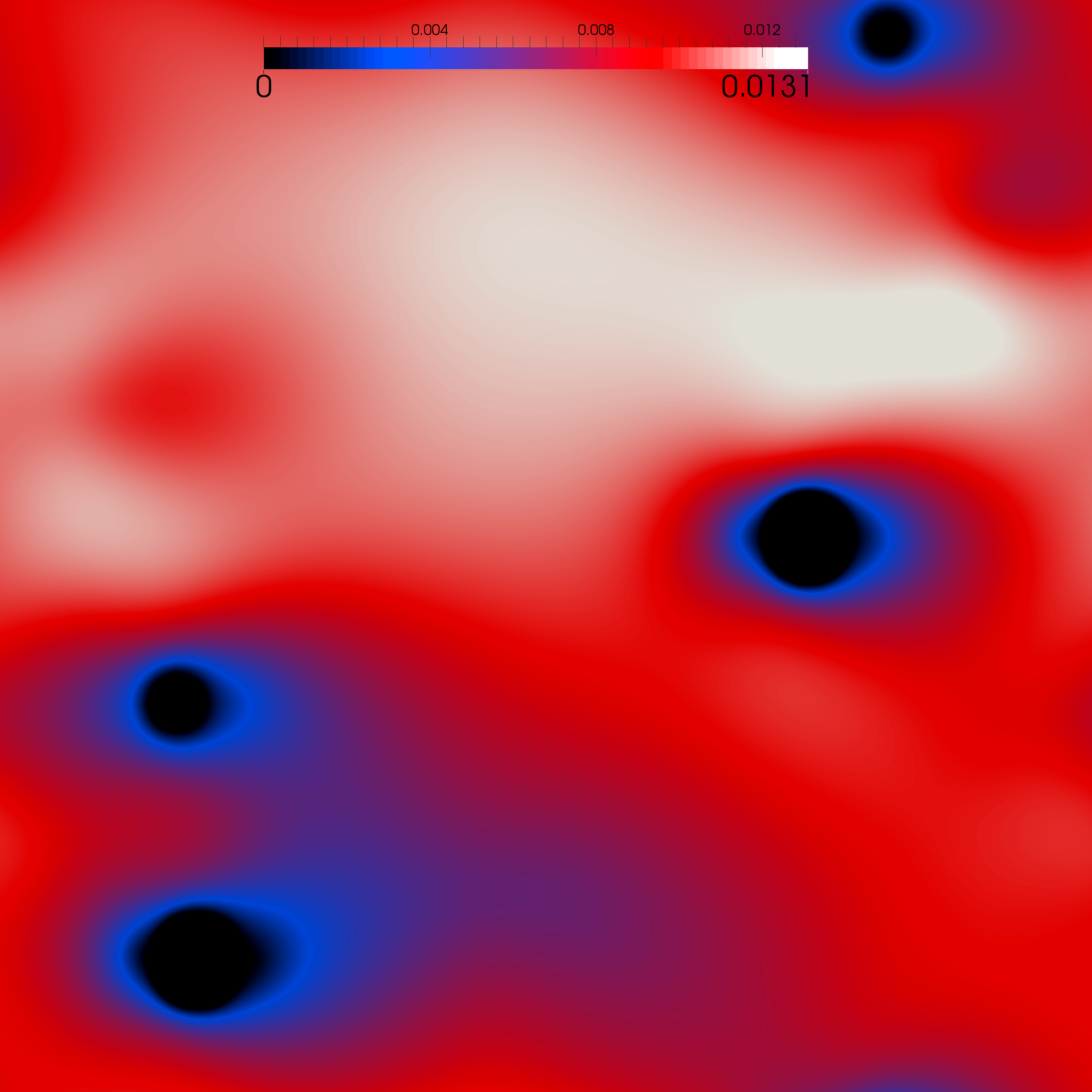}
\includegraphics[width=0.5\linewidth]{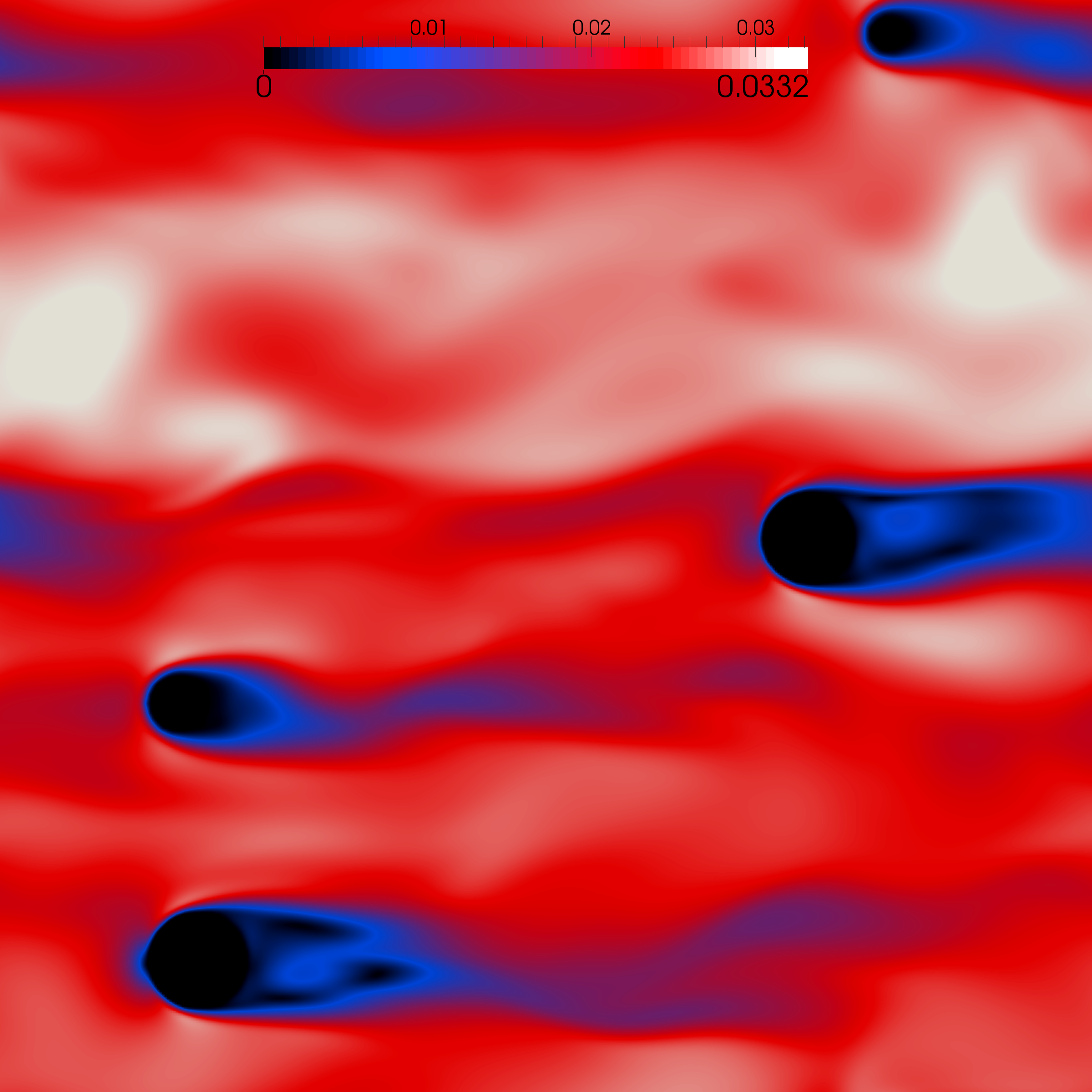}
\caption{Velocity magnitude plot (slice $y=50$) in particle bed of volume fraction
  $\varphi=0.01$ in the laminar regime ($Re_p = 2.4$, left), and in the transient regime
  ($Re_p = 300$, right). Flow direction is from left two right in both cases. As indicated
  by the color bar, the flow velocity is zero at the black spots where the particles are
  located, and highest within the white areas.}
\label{fig:velocity}
\end{figure*}

Fig.~\ref{fig:streamlines} shows the flow profile of fluid through a bed of spheres with
solid volume fraction $\varphi=0.35$ at a low Reynolds number compared with the profile at a
high Reynolds number $Re_p = 300$ through the same bed. To visualize the flow, a number of
streamlines has been traced through the flow field, starting at equally spaced points
along a diagonal line in the back plane, $x=0$, of the domain with the flow directed along
the x-axis. The plots differ significantly at the two different regimes.

As a second example, we consider the velocity field in a bed of low volume fraction
$\varphi = 0.01$. Fig.~\ref{fig:velocity} shows the velocity magnitude, indicated by
color, in a plane slicing through the domain, at two different regimes. In the transient
case ($Re_p = 300$) we observe the appearance of eddies behind the particles that are
located at the black spots.

\subsection{Study: Drag force within random static assemblies of spheres}
\label{sec:comprehensive}
\begin{figure}
  \centering
  \includegraphics[width=0.55\linewidth]{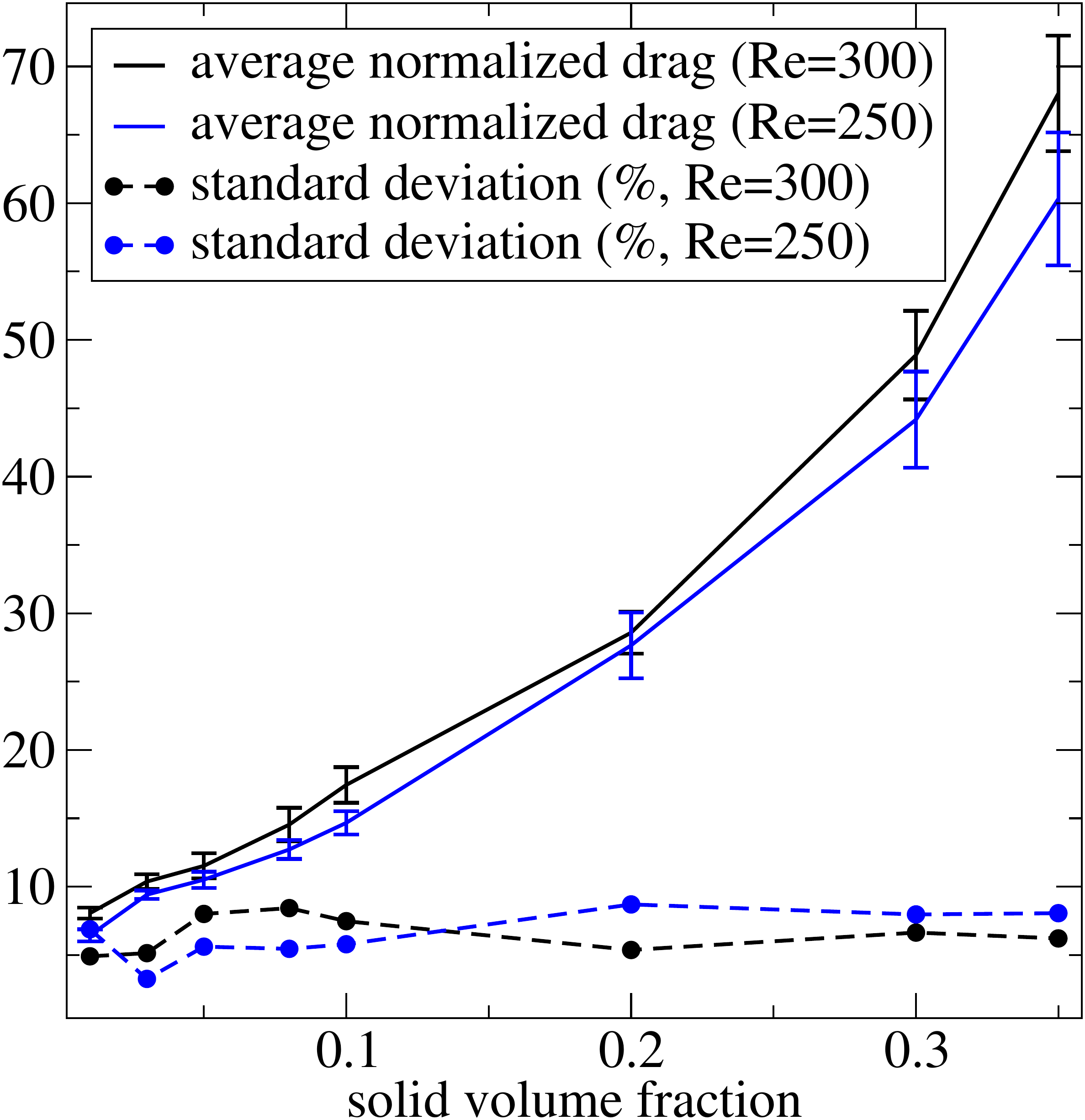}
  \caption{Normalized drag values for $Re_p=250,\,300$. Each data point corresponds to the mean value of $5$
    runs. The standard deviation is shown as error bars and separately in percent. The
    percent values are strictly below $10\%$ and not correlated with the solid volume
    fraction.}
  \label{fig:deviation}
\end{figure}
\begin{figure}
  \centering
  \includegraphics[width=0.7\linewidth]{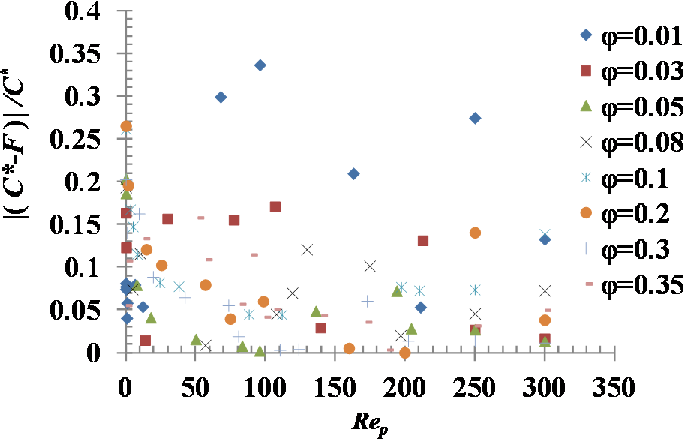}
  \caption{Pointwise relative deviation of the computed values $C^*$ from the fit
    correlation $F$ given by Eq.~(\ref{eq:drag}).}
  \label{fig:error}
\end{figure}
For the present study of the drag force within mono-disperse fixed assemblies of spheres,
the solid volume fraction takes the values $\varphi \in$ $\{0.01,$ $0.03,$ $0.05,$ $0.08,$
$0.1,$ $0.2,$ $0.3,$ $0.35 \}$, while the particle Reynolds number $Re_p$ is varied from
$0.05$ to $300$.  A series of simulations was carried out, repeating each pair $(\varphi,
Re_p)$ at least five times.  In our approach, the average flow velocity in the domain may
vary depending on the randomly generated bed. Hence, the resulting Reynolds numbers has
also been averaged for each series. For higher target Reynolds numbers ($Re_p \geq 200$),
however, the gravitation parameter $g$ was adjusted dynamically to fix the Reynolds
number. The resolution in this study was varied between $d=17$ and $d=47$ lattice units
per sphere diameter, and the number of spheres taken for each simulation was either $N=27$
or $N=54$. A cubic domain of length varying from 74 to 449 lattice units, depending on the
desired volume fraction $\varphi$, was used with periodic boundary conditions for all the
exterior boundaries. The domain length to particle diameter ratio varied between $3.43$
and $11.18$.  For each run a randomized arrangement of spheres is generated, and the
resulting dimensionless drag values are averaged over the whole series to obtain the
approximated bed-independent dimensionless drag force. The number of 5 runs was found to
be sufficient to even out artifacts from the random bed generation, in consistency with
the reports by \citet{Hill2001a} and \citet{Tenneti2011}. We did not apply any
``check-reject-repeat'' -- strategy to eliminate any spikes in our data. The variation of the
resulting absolute values is largest for higher $Re_p$. Fig.~\ref{fig:deviation} shows the
data for the two highest Reynolds numbers included in this study, and the standard
deviation for each data point. The situation is similar for the lower Reynolds number,
however with smaller absolute values.

A dimensionless drag correlation $F$ is obtained using the conjugate gradient
method to minimize the objective function
\begin{equation*} 
  E = \sum_{(\varphi, Re_p)} \frac{|F - C^{*}|}{F},
\end{equation*}
between the correlation and the LBM simulated values $C^{*}$ (averaged over $5$ random
beds) with an average absolute percentage error of $ 9.69 \%$ compared with the simulated
data and with a coefficient of determination $R^2 = 0.995$. Using the model function
$F(\varphi, Re_p) = (1-\varphi)^{a} [ b + c {Re_p}^d + e (1+Re_p)^{-f \varphi} + g
(1+Re_p)^{-h \varphi} ]$, with parameters $a, b, .., h$, one obtains the correlation $C =
F(\varphi, Re_p)$, with
\begin{equation}
\begin{split}
F = & (1-\varphi)^{-5.726} \big[ 1.751 + 0.151 \, {Re_p}^{0.684} \\
& - 0.445 \, (1+Re_p)^{1.04 \varphi} - 0.16 \, (1+Re_p)^{0.0003 \varphi} \big] .
\label{eq:drag}
\end{split}
\end{equation}

Fig.~\ref{fig:error} shows the pointwise deviation $|F - C^{*}|/C^{*}$ between Eq.~(\ref{eq:drag}) and the
simulated values. One observes the largest deviations for small volume
fractions.
The average absolute percentage error with
respect to $C^{*}$ is $9.3 \%$. Unlike the correlations reported by other authors
\citep[for instance][]{Beetstra2007,Benyahia2006}, which consist of two parts; the first
part for the Stokes region, i.e., $Re_p \ll 1$, and a function of volume fraction only;
the second part a function of solid volume fraction and $Re_p$, we chose a unified
expression.

\begin{figure}
  \centering
  \begin{subfigure}[t]{0.48\linewidth}
    \includegraphics[width=\textwidth]{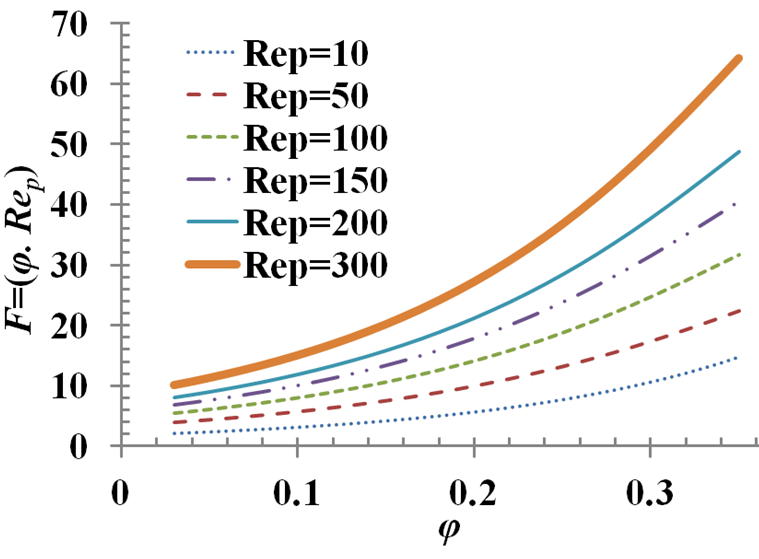}
    \caption{Normalized drag force based on total hydrodynamic force as a function of
      solid volume fraction for different particle Reynolds number. }
    \label{fig:Rep}
  \end{subfigure}\hspace{0.02\linewidth}
  \begin{subfigure}[t]{0.48\linewidth}
  \includegraphics[width=\textwidth]{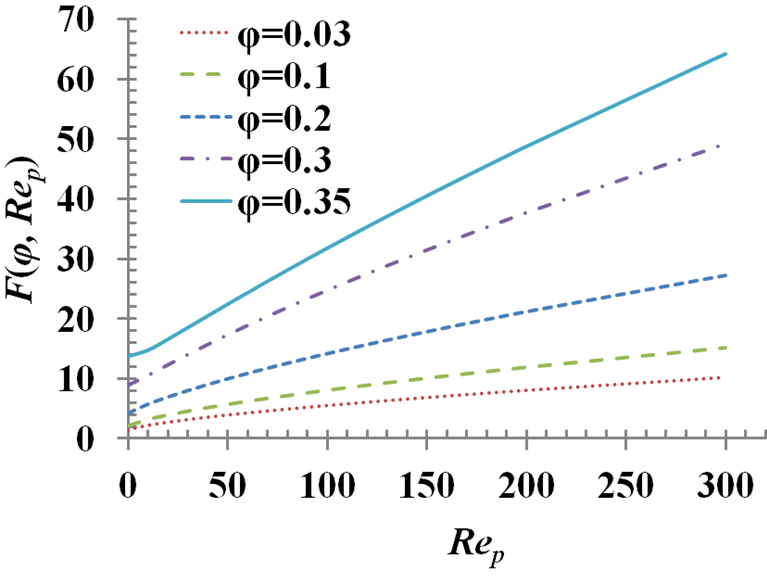}
  \caption{The normalized drag force as a function of the Reynolds number $Re_p$ for various
    solid volume fractions.}
  \label{fig:phi}
  \end{subfigure}

  \caption{Drag correlation derived from present study.}
\end{figure}

Fig.~\ref{fig:Rep} shows the effect of $\varphi$ on the normalized drag force at constant
Reynolds number. It can be seen that the increase in drag force with increase in $\varphi$
is more significant at high Reynolds number. Similarly, it can be seen from
Fig.~\ref{fig:phi} that at a constant $\varphi$ the normalized drag force increases with
the Reynolds number. The slope becomes steeper with increasing concentration $\varphi$. At
low concentrations and Reynolds number beyond $Re_p>200$, the increase of drag with $Re_p$
is least significant.

\subsection{Comparison with previous studies}
\label{sec:comparison}
\begin{figure}
  \centering
  \includegraphics[width=0.72\linewidth]{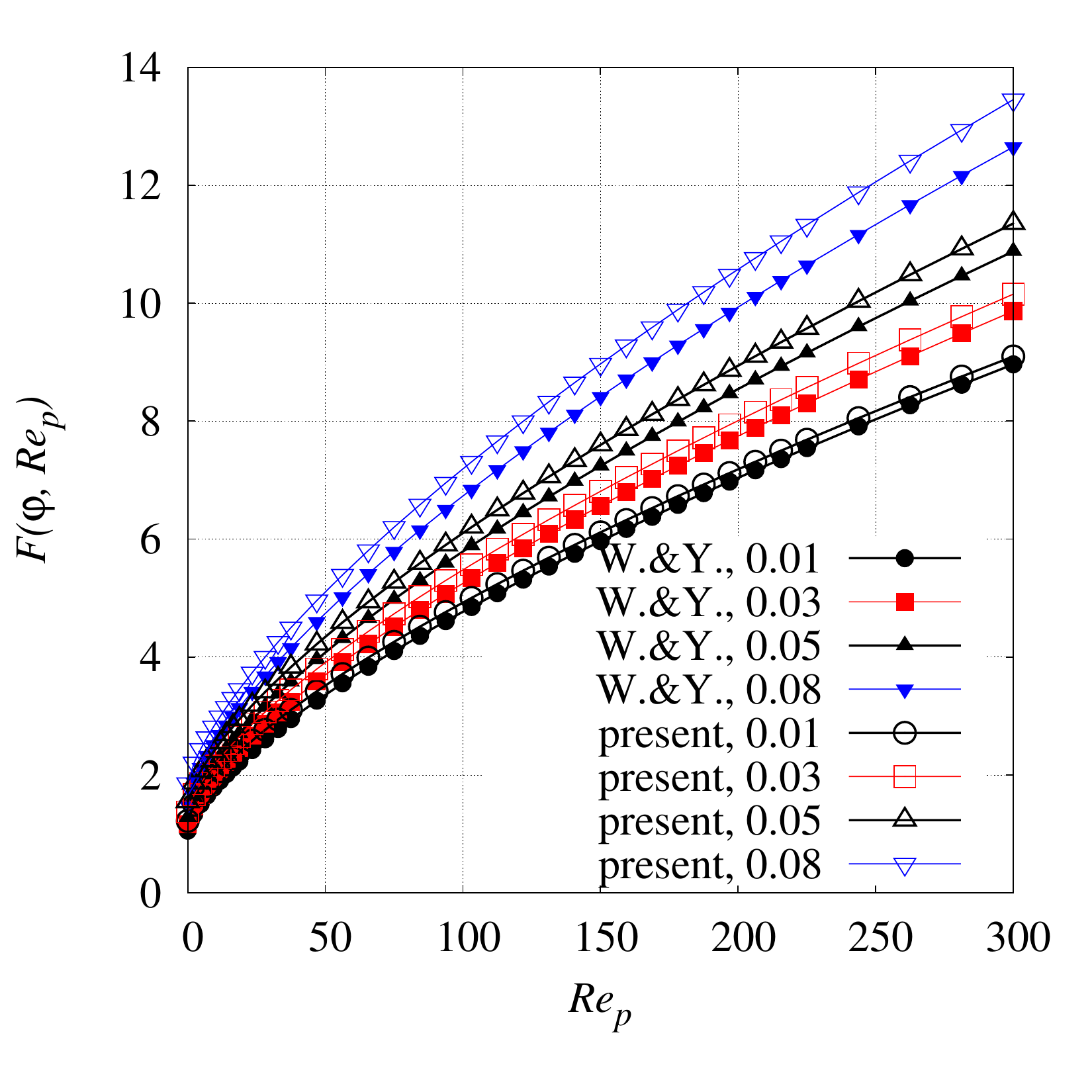}
  \caption{Wen \& Yu - correlation for low solid volume fractions $\varphi=0.01 \,..\, 0.08$
    (filled symbols) in comparison to the present study (hollow symbols).}
  \label{fig:WenYu}

\end{figure}

\begin{figure}
  \centering

  \begin{subfigure}[t]{0.49\linewidth}
    \includegraphics[width=\linewidth]{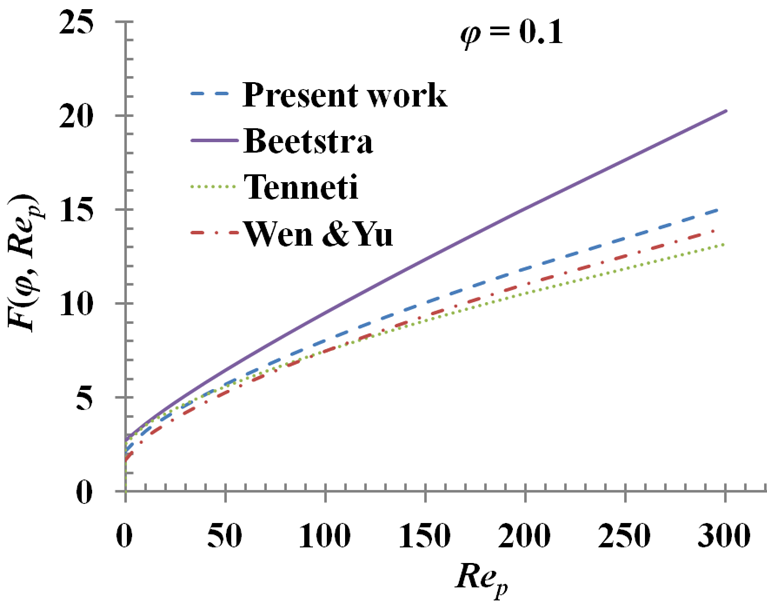}
  \end{subfigure}
  \begin{subfigure}[t]{0.49\linewidth}
    \includegraphics[width=\linewidth]{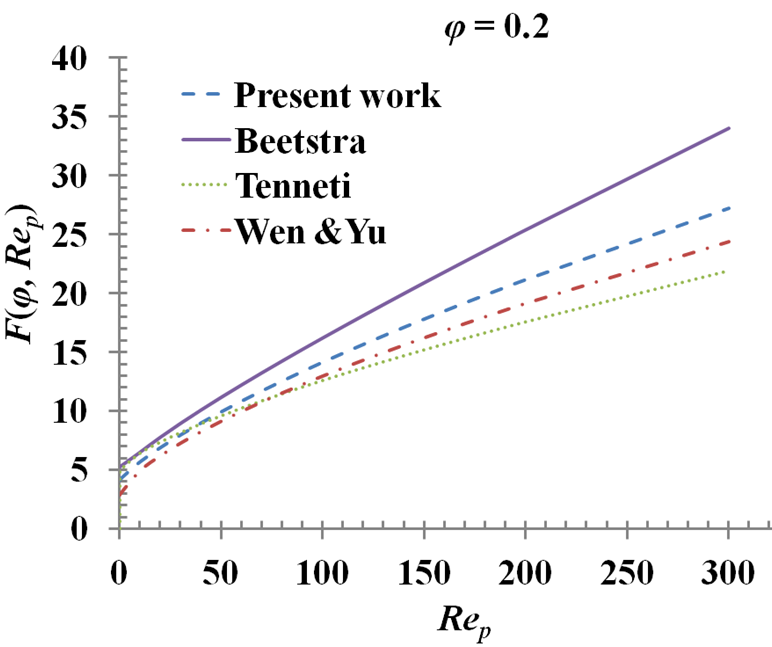}
  \end{subfigure}
  \begin{subfigure}[b]{0.49\linewidth}
    \includegraphics[width=\linewidth]{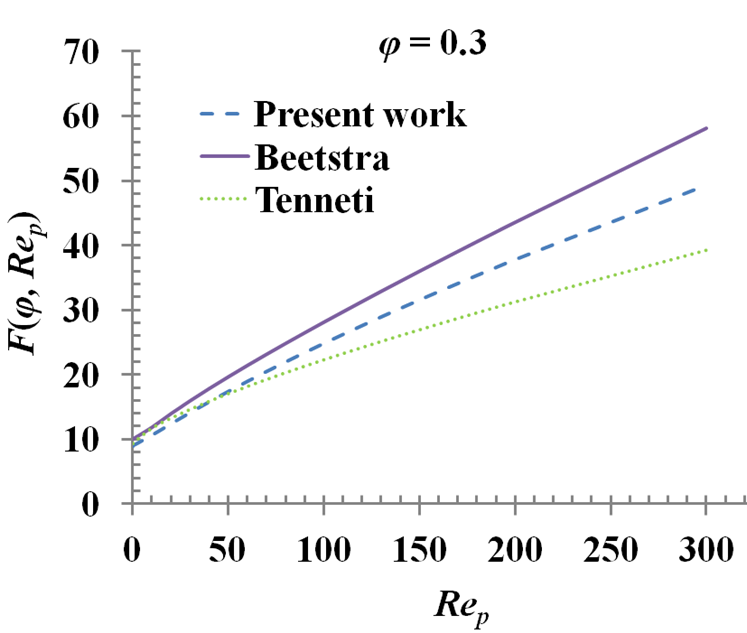}
  \end{subfigure}
  \begin{subfigure}[b]{0.49\linewidth}
    \includegraphics[width=\linewidth]{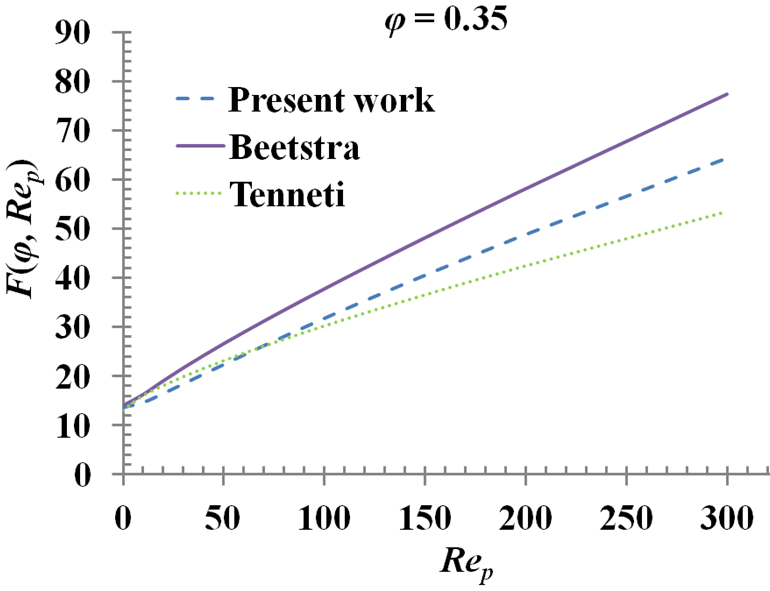}
  \end{subfigure}
  
  \caption{Normalized drag force predicted by our correlation compared to that reported by
    other authors.}
  \label{fig:comparison}
\end{figure}

At low solid volume fractions $\varphi \lesssim 0.2$, there is the well-established
correlation by \citet{Wen1966}, compiled from experimental data. Because of the emphasis
on the dilute case in our study, we compare to this relation first. Fig.~\ref{fig:WenYu}
shows that the newly presented correlation is close to the Wen \& Yu - relation for the
most dilute cases included in this study. The smallest average deviation is $+4.3 \%$ for
$\varphi = 0.01$, and increases with $\varphi$. The average deviation for $\varphi=0.2$ is
$+10.8 \%$ as shown in Fig.~\ref{fig:comparison}. Note, that the correlation given by Wen
and Yu is applicable only for $\varphi \lesssim 0.2$, whereas the proposed correlation can
be used up to $\varphi \approx 0.35$.

Fig.~\ref{fig:comparison} shows the drag law predicted by Eq. (\ref{eq:drag}) in
comparison to the correlations obtained in the numerical studies by \citet{Beetstra2007}
and \citet{Tenneti2011}. In these references, both the simulated hydrodynamic force and
average flow rate is averaged over all runs of each $(\varphi, Re_p)$ -- pair, to
therefrom obtain the unknown dimensionless drag. The maximum and minimum average deviation
when compared with the drag law reported by \citet{Tenneti2011} is approximately $+12 \%$
for $\varphi = 0.2$, and $+7 \%$ for $\varphi = 0.35$, respectively. The deviation
increases with the Reynolds number. When compared with the results reported by \citet
{Beetstra2007}, the maximum and minimum average deviation is approximately $-20\%$ for
$\varphi = 0.1$ and $-14\%$ for $\varphi =0.3$, respectively. \citet{Tenneti2011} comment
that the deviation of more than $30\%$ between the values reported by \citet
{Beetstra2007} and their own is due to the poor lattice resolutions applied in the latter
work. Hence, of the two references, the study of \citet{Tenneti2011} should be regarded
more accurate. The lattice resolutions for the present study are comparable those applied
by Tenneti, and have been chosen based on the considerations of
Sec.~\ref{sec:resolution}. The present LBM-scheme does not suffer from a viscosity
dependent error at the solid boundaries. As we have also demonstrated numerically
(cf. Tab.~\ref{tab:magic}) for $Re=300$, where the deviation from the other studies was
most significant, this deviation is independent from the parameterization according to
Eq.~(\ref{eq:magic}). The sensitivity study of Sec.~\ref{sec:resolution} indicates that
errors due to unresolved flow phenomena do not affect the present study
significantly. However, a small error stemming from the first-order staircase
approximation of the sphere boundaries is present and can roughly be estimated to yield an
overshoot of a few percent. We believe that the deviations between the present study and
that of \citet{Tenneti2011} is partially due to that reason.


\section{Conclusion}
Fluid flow through random static assemblies of spheres has been studied by means of the
lattice Boltzmann method. Compared with previous studies, our approach does not suffer
from viscosity dependent boundary conditions. As a result, a wide range of Reynolds
numbers can be studied by simple adjustment of the lattice viscosity - provided that a
sufficient grid resolution is chosen. A convergence study demonstrates that the results
are effectively independent of the lattice resolution. From systematic numerical studies
of flow through random arrays at varying solid volume fraction ($0.01 \leq \varphi \leq
0.35$), a new correlation for the average normalized drag force for particle Reynolds
numbers up to $300$ is obtained.

Despite a certain expected error from the first order boundary conditions (``bounce
back-rule''), we successfully showed the reliability of the present approach for the given
range of moderate Reynolds numbers. The comparison with \citet{Wen1966} shows that the
correlation predicts well at very low concentrations ($\varphi < 0.1$) and at the same
time can be used up to a solid volume fraction of $0.35$. The deviations of the results
obtained with our lattice Boltzmann scheme to the simulations of \citet{Tenneti2011} are
not as dramatic as reported in the same reference in comparison to previous lattice
Boltzmann studies. This confirms the reliability of the LBM. On the other hand, the
existing deviations between different numerical approaches also indicate a need for
further studies conducted with more accurate numerical schemes in the future, as it is
presently unclear which results are the more reliable. This could be achieved by using a
lattice Boltzmann method equipped with a higher order boundary scheme
\citep{Bouzidi2001,Ginzburg2003,MeiYuShyyLuo2002}, for instance.

The normalized drag correlation proposed in the present study will be useful in predicting
the flow behavior of particles in fluid-particulate systems, where the solid volume
fractions range from very low to moderate values. Usually, more than one correlations are
used when the range of solid volume fraction and Reynolds number is large. The correlation
proposed by us has a simpler expression compared to those proposed by, e.g.,
\citet{Beetstra2007} or \citet{Benyahia2006}, and is purely based on LBM simulated
data. Future work will aim at obtaining a correlation that can be used for higher solid
volume fractions and higher Reynolds numbers.


\section*{Acknowledgements}
The first author would like to thank the Bayerische Forschungsstiftung and KONWIHR project
waLBerla-EXA for financial support. The second author, S. Mohanty, would like to
acknowledge the Alexander von Humboldt Foundation, Bonn, Germany, for the financial
support and also the Director, CSIR-Institute of Minerals \& Materials Technology,
Bhubaneswar, for permission to publish this paper.  Further, we would like to thank the
Regionales Rechenzentrum Erlangen (RRZE) for providing us with the computational
resources.

\bibliographystyle{plainnat}
\bibliography{lit}

\begin{thebibliography}{45}
\providecommand{\natexlab}[1]{#1}
\providecommand{\url}[1]{\texttt{#1}}
\expandafter\ifx\csname urlstyle\endcsname\relax
  \providecommand{\doi}[1]{doi: #1}\else
  \providecommand{\doi}{doi: \begingroup \urlstyle{rm}\Url}\fi

\bibitem[Aidun and Clausen(2010)]{AidunClausen}
C.~K. Aidun and J.~R. Clausen.
\newblock Lattice-{B}oltzmann method for complex flows.
\newblock \emph{Annu. Rev. Fluid Mech.}, 42:\penalty0 439--472, 2010.

\bibitem[Beetstra et~al.(2007)Beetstra, van~der Hoef, and
  Kuipers]{Beetstra2007}
R.~Beetstra, M.~A. van~der Hoef, and J.~A.~M. Kuipers.
\newblock Drag force of intermediate {R}eynolds number flow past mono- and
  bidisperse arrays of spheres.
\newblock \emph{Fluid Mechanics and Transport Phenomena}, 53(2):\penalty0
  489--501, 2007.
\newblock \doi{10.1002/aic.11065}.

\bibitem[Benyahia et~al.(2006)Benyahia, Syamlal, and O'Brien]{Benyahia2006}
S.~Benyahia, M.~Syamlal, and T.~J. O'Brien.
\newblock Extension of {Hill-Koch-Ladd} - drag correlation over all ranges of
  {R}eynolds number and solids volume fraction.
\newblock \emph{Powder Technology}, 162:\penalty0 166--174, 2006.

\bibitem[Benzi et~al.(1992)Benzi, Succi, and Vergassola]{BenziEtAl}
R.~Benzi, S.~Succi, and M.~Vergassola.
\newblock The lattice {B}oltzmann equation: theory and applications.
\newblock \emph{Physics Reports}, 222(3):\penalty0 145--197, 1992.

\bibitem[Bogner and R{\"u}de(2013)]{Bogner2013}
S.~Bogner and U.~R{\"u}de.
\newblock Simulation of floating bodies with the lattice {B}oltzmann method.
\newblock \emph{Computers and Mathematics with Applications}, 65:\penalty0
  901--913, 2013.

\bibitem[Bouzidi et~al.(2001)Bouzidi, Firdaouss, and Lallemand]{Bouzidi2001}
M.~Bouzidi, M.~Firdaouss, and P.~Lallemand.
\newblock Momentum transfer of a {B}oltzmann-lattice fluid with boundaries.
\newblock \emph{Physics of Fluids}, 13(11):\penalty0 3452, 2001.

\bibitem[Buick and Greated(2000)]{BuickGreated2000}
J.M. Buick and C.A. Greated.
\newblock Gravity in a lattice {B}oltzmann model.
\newblock \emph{Phys. Rev. E}, 61(5):\penalty0 5307, 2000.

\bibitem[Chen and Doolen(1998)]{ChenDoolen1998}
S.~Chen and G.~D. Doolen.
\newblock Lattice {B}oltzmann method for fluid flows.
\newblock \emph{Annu. Rev. Fluid Mech.}, 30:\penalty0 329--364, 1998.

\bibitem[Chu et~al.(2009)Chu, Wang, Yu, Vince, Barnett, and Barnett]{Chu2009}
K.~W. Chu, B.~Wang, A.~B. Yu, A.~Vince, G.~D. Barnett, and P.~J. Barnett.
\newblock {CFD--DEM} study of the effect of particle density distribution on
  the multiphase flow and performance of dense medium cyclone.
\newblock \emph{Minerals Engineering}, 22:\penalty0 893--909, 2009.

\bibitem[d'Humieres and Ginzburg(2009)]{Ginzburg2009}
D.~d'Humieres and I.~Ginzburg.
\newblock Viscosity independent numerical errors for lattice {B}oltzmann
  models: From recurrence equations to {"}magic{"} collision numbers.
\newblock \emph{Computers and Mathematics with Applications}, 58:\penalty0
  823--840, 2009.

\bibitem[Ergun(1952)]{Ergun1952}
S.~Ergun.
\newblock Fluid flow through packed columns.
\newblock \emph{Chemical Engineering Progress}, 48:\penalty0 89--94, 1952.

\bibitem[Feichtinger et~al.(2011)Feichtinger, Donath, K{\"o}stler, G{\"o}tz,
  and R{\"u}de]{walberla2011}
C.~Feichtinger, S.~Donath, H.~K{\"o}stler, J.~G{\"o}tz, and U.~R{\"u}de.
\newblock Wa{LB}erla: {HPC} software design for computational engineering
  simulations.
\newblock \emph{Journal of Computational Science}, 2(2):\penalty0 105--112,
  2011.
\newblock \doi{10.1016/j.jocs.2011.01.004}.

\bibitem[Gidaspow(1986)]{Gidaspow1986}
D.~Gidaspow.
\newblock Hydrodynamics of fluidization and heat transfer: supercomputer
  modeling.
\newblock \emph{Applied Mechanics Review}, 39:\penalty0 1--23, 1986.

\bibitem[Ginzbourg and Adler(1994)]{Ginzburg1994}
I.~Ginzbourg and P.M. Adler.
\newblock Boundary flow condition analysis for three-dimensional lattice
  {B}oltzmann model.
\newblock \emph{Journal of Physics II France}, 4:\penalty0 191--214, 1994.

\bibitem[Ginzburg(2005)]{Ginzburg2005}
I.~Ginzburg.
\newblock Equilibrium-type and link-type lattice {B}oltzmann models for generic
  advection and anisotropic-dispersion equation.
\newblock \emph{Advances in Water Resources}, 28:\penalty0 1171--1195, 2005.

\bibitem[Ginzburg and d'Humieres(2003)]{Ginzburg2003}
I.~Ginzburg and D.~d'Humieres.
\newblock Multireflection boundary conditions for lattice {B}oltzmann models.
\newblock \emph{Physical Review E}, 68:\penalty0 066614--1 -- 066614--29, 2003.

\bibitem[Ginzburg et~al.(2008)Ginzburg, Verhaeghe, and
  d'Humieres]{Ginzburg2007}
I.~Ginzburg, F.~Verhaeghe, and D.~d'Humieres.
\newblock Two-relaxation-time lattice {B}oltzmann scheme: About
  parametrization, velocity, pressure and mixed boundary conditions.
\newblock \emph{Communications in Computational Physics}, 3(2):\penalty0
  427--478, 2008.

\bibitem[G{\"o}tz et~al.(2010)G{\"o}tz, Iglberger, St{\"u}rmer, and
  R{\"u}de]{Goetz2010}
J.~G{\"o}tz, K.~Iglberger, M.~St{\"u}rmer, and U.~R{\"u}de.
\newblock Direct numerical simulation of particulate flows on 294912 processor
  cores.
\newblock In \emph{Proceedings of the 2010 ACM/IEEE International Conference
  for High Performance Computing, Networking, Storage and Analysis}, pages
  1--11. IEEE Computer Society, 2010.

\bibitem[Guo et~al.(2002)Guo, Zheng, and Shi]{Guo2002}
Z.~Guo, Chuguang Zheng, and Baochang Shi.
\newblock Discrete lattice effects on the forcing term in the lattice
  {B}oltzmann method.
\newblock \emph{PHYSICAL REVIEW E}, 65:\penalty0 046308--1 -- 046308--6, 2002.

\bibitem[Hasimoto(1959)]{Hasimoto1959}
H.~Hasimoto.
\newblock On the periodic fundamental solutions of the {S}tokes equations and
  their application to viscous flow past a cubic array of spheres.
\newblock \emph{Journal of Fluid Mechanics}, 5:\penalty0 317--328, 1959.

\bibitem[Hill et~al.(2001{\natexlab{a}})Hill, Koch, and Ladd]{Hill2001a}
R.~J. Hill, D.~L. Koch, and A.~J.~C. Ladd.
\newblock The first effects of fluid inertia on flows in ordered and random
  arrays of spheres.
\newblock \emph{Journal of Fluid Mechanics}, 448:\penalty0 213--241,
  2001{\natexlab{a}}.
\newblock \doi{10.1017/S0022112001005948}.

\bibitem[Hill et~al.(2001{\natexlab{b}})Hill, Koch, and Ladd]{Hill2001b}
R.~J. Hill, D.~L. Koch, and A.~J.~C. Ladd.
\newblock Moderate-{R}eynolds-number flows in ordered and random arrays of
  spheres.
\newblock \emph{Journal of Fluid Mechanics}, 448:\penalty0 243--278,
  2001{\natexlab{b}}.
\newblock \doi{10.1017/S0022112001005936}.

\bibitem[Holloway et~al.(2010)Holloway, Yin, and Sundaresan]{Holloway2010}
W.~Holloway, X.~Yin, and S.~Sundaresan.
\newblock Fluid-particle drag in inertial polydisperse gas-solid suspensions.
\newblock \emph{American Institute of Chemical Engineers Journal},
  56(8):\penalty0 1995--2004, 2010.

\bibitem[Junk and Yang(2005)]{JunkYang2005}
M.~Junk and Z.~Yang.
\newblock Asymptotic analysis of lattice boltzmann boundary conditions.
\newblock \emph{Journal of Statistical Physics}, 121 (1/2):\penalty0 3--35,
  2005.

\bibitem[{Ladd}(1994{\natexlab{a}})]{Ladd1994a}
A.-J.-C. {Ladd}.
\newblock Numerical simulations of particulate suspensions via a discretized
  {B}oltzmann equation. {P}art 1. {T}heoretical foundation.
\newblock \emph{Journal of Fluid Mechanics}, 271:\penalty0 285--309, July
  1994{\natexlab{a}}.

\bibitem[{Ladd}(1994{\natexlab{b}})]{Ladd1994b}
A.-J.-C. {Ladd}.
\newblock Numerical simulations of particulate suspensions via a discretized
  {B}oltzmann equation. {P}art 2. {N}umerical results.
\newblock \emph{Journal of Fluid Mechanics}, 271:\penalty0 311--339, July
  1994{\natexlab{b}}.

\bibitem[Ladd and Verberg(2001)]{Ladd2001}
A.~J.~C. Ladd and R.~Verberg.
\newblock Lattice-{B}oltzmann simulations of particle-fluid suspensions.
\newblock \emph{Journal of Statistical Physics}, 104:\penalty0 1191--1251,
  2001.

\bibitem[Luo(1998)]{Luo1998}
L.-S. Luo.
\newblock Unified theory of lattice {B}oltzmann models for nonideal gases.
\newblock \emph{Phys. Rev. Lett.}, 81(8):\penalty0 1618, 1998.

\bibitem[Mei et~al.(2002)Mei, Yu, Shyy, and Luo]{MeiYuShyyLuo2002}
R.~Mei, D.~Yu, W.~Shyy, and L.-S. Luo.
\newblock Force evaluation in the lattice {B}oltzmann method involving curved
  geometry.
\newblock \emph{Phys. Rev. E}, 65:\penalty0 041203, Apr 2002.

\bibitem[Mishra and Tripathy(2010)]{Mishra2010}
B.~K. Mishra and A.~Tripathy.
\newblock Preliminary study of particle separation in spiral concentrators
  using dem.
\newblock \emph{International Journal of Mineral Processing}, 94:\penalty0
  192--195, 2010.

\bibitem[Mohanty et~al.(2011)Mohanty, Das, and Mishra]{Mohanty2011}
S.~Mohanty, B.~Das, and B.K. Mishra.
\newblock A preliminary investigation into magnetic separation process using
  cfd.
\newblock \emph{Minerals Engineering}, 24:\penalty0 1651--1657, 2011.

\bibitem[Pan et~al.(2006)Pan, Luo, and Miller]{Pan2005}
C.~Pan, L.-S. Luo, and C.~T. Miller.
\newblock An evaluation of lattice {B}oltzmann schemes for porous medium flow
  simulation.
\newblock \emph{Computers \& Fluids}, 25:\penalty0 898--909, 2006.

\bibitem[Qian et~al.(1992)Qian, d'Humieres, and Lallemand]{QianEtAl1992}
Y.~H. Qian, D.~d'Humieres, and P.~Lallemand.
\newblock Lattice {B}{G}{K} models for {N}avier-{S}tokes equations.
\newblock \emph{Europhysical Letters}, 17(6):\penalty0 479--484, 1992.

\bibitem[Richardson and Zaki(1954)]{Richardson1954}
J.~F. Richardson and W.~N. Zaki.
\newblock Sedimentation and fluidisation. part i.
\newblock \emph{Transactions of the Institution of Chemical Engineers},
  32:\penalty0 35--53, 1954.

\bibitem[Sangani and Acrivos(1982)]{Sangani1982}
A.~S. Sangani and A.~Acrivos.
\newblock Slow flow through a periodic array of spheres.
\newblock \emph{International Journal of Multiphase Flow}, 8(4):\penalty0
  343--360, 1982.

\bibitem[Shyamlal and O'Brien(1987)]{Shyamlal1987}
M.~Shyamlal and T.~J. O'Brien.
\newblock A generalized drag correlation for multiparticle systems.
\newblock \emph{Morgantown Energy Technology Center DOE Report}, 1987.

\bibitem[Succi(2001)]{Succi}
S.~Succi.
\newblock \emph{The Lattice {B}oltzmann Equation for Fluid Dynamics and
  Beyond}.
\newblock Oxford Science Publications, 2001.

\bibitem[Swain and Mohanty(2013)]{Swain2013}
S.~Swain and S.~Mohanty.
\newblock A 3-dimensional {E}ulerian-{E}ulerian {CFD} simulation of a
  hydrocyclone.
\newblock \emph{Applied Mathematical Modelling}, 37:\penalty0 2921--2932, 2013.

\bibitem[Tenneti et~al.(2011)Tenneti, Garg, and Subramaniam]{Tenneti2011}
S.~Tenneti, R.~Garg, and S.~Subramaniam.
\newblock Drag law for monodisperse gas-solid systems using particle-resolved
  direct numerical simulation of flow past fixed assemblies of spheres.
\newblock \emph{International Journal of Multiphase Flow}, 37:\penalty0
  1072--1092, 2011.

\bibitem[van~der Hoef et~al.(2005)van~der Hoef, Beetstra, and
  Kuipers]{vanDerHoef2005}
M.~A. van~der Hoef, R.~Beetstra, and J.~A.~M. Kuipers.
\newblock Lattice-{B}oltzmann simulations of low-{R}eynolds-number flow past
  mono- and bidisperse arrays of spheres: results for the permeability and drag
  force.
\newblock \emph{Journal of Fluid Mechanics}, 528:\penalty0 233--254, 2005.
\newblock \doi{10.1017/S0022112004003295}.

\bibitem[van~der Hoef et~al.(2006)van~der Hoef, Ye, van Sint~Annaland, Andrews,
  Sundaresan, and Kuipers]{Hoeff2006}
M.~A. van~der Hoef, M.~Ye, M.~van Sint~Annaland, A.~T. Andrews, S.~Sundaresan,
  and J.~A.~M. Kuipers.
\newblock Multiscale modeling of gas-fluidized beds.
\newblock \emph{Advances in Chemical Engineering}, 31:\penalty0 64--149, 2006.

\bibitem[Wen and Yu(1966)]{Wen1966}
C.~Y. Wen and Y.~H. Yu.
\newblock Mechanics of fluidization.
\newblock \emph{Chem Eng Prog Symp Ser.}, 62:\penalty0 100--111, 1966.

\bibitem[Yin and Sundaresan(2009{\natexlab{a}})]{Yin2009a}
X.~Yin and S.~Sundaresan.
\newblock Drag law for bidisperse gas-solid suspensions containing equally
  sized spheres.
\newblock \emph{Industrial and Engineering Chemistry Research}, 48:\penalty0
  227--241, 2009{\natexlab{a}}.

\bibitem[Yin and Sundaresan(2009{\natexlab{b}})]{Yin2009b}
X.~Yin and S.~Sundaresan.
\newblock Fluid-particle drag in low-{R}eynolds-number polydisperse gas-solid
  suspensions.
\newblock \emph{American Institute of Chemical Engineers Journal}, 55:\penalty0
  1352--1368, 2009{\natexlab{b}}.

\bibitem[Yu et~al.(2003)Yu, Mei, Luo, and Shyy]{Yu2003}
D.~Yu, R.~Mei, L.-S. Luo, and W.~Shyy.
\newblock Viscous flow computations with the method of lattice {B}oltzmann
  equation.
\newblock \emph{Progress in Aerospace Sciences}, 39:\penalty0 329--367, 2003.

\end{thebibliography}

\end{document}